\documentclass[11pt]{article}
\usepackage[table,xcdraw]{xcolor}
\usepackage[utf8]{inputenc}
\usepackage[margin=2.5cm]{geometry}
\usepackage{graphicx}
\usepackage{setspace}
\usepackage{amsmath}
\usepackage{amssymb}
\usepackage{url}
\usepackage{longtable}
\usepackage{amssymb}
\usepackage{pdflscape}
\usepackage{hyperref}
\usepackage{xcolor}
\usepackage{tikz}
\usepackage{xcolor}
\usepackage{tabularx}
\usepackage{booktabs}
\usepackage{makecell}
\usepackage{titlesec}
\usepackage[defaultlines=2,all]{nowidow}
\usepackage{caption}
\usepackage{hyperref}
\usepackage{nameref}
\titlespacing{\paragraph}{0pt}{0pt}{1ex}
\usepackage[margin=2cm,font=footnotesize]{caption}
\usepackage{dcolumn}
\usepackage{rotating}
\usepackage{tikz}

\usepackage[export]{adjustbox}

\usepackage[backend=biber,style=authoryear,
citestyle=authoryear,isbn=false,url=false,eprint=false,
giveninits=true,maxcitenames=1,uniquename=init]{biblatex}
\AtEveryBibitem{
  \clearfield{note}
  \clearfield{url}
  \clearfield{issn}
  \clearfield{urldate}
  \clearfield{month}
}

\definecolor{bleucite}{RGB}{34,111,212}
\usepackage{hyperref}
\hypersetup{hidelinks}
\AtEveryCite{\color{bleucite}}

\usepackage{cleveref}

\renewcommand{\epsilon}{\varepsilon}

\renewcommand{\t}[1]{\texttt{#1}}

\title{Connectivity and Community Structure of Online and Register-based Social Networks }

\author{Márton Menyhért$^1$, Eszter Bokányi$^{2,*}$, Rense Corten$^3$, Eelke M. Heemskerk$^2$,\\[0.3ex]
Yuliia Kazmina$^2$, Frank W. Takes$^1$}

\date{{\footnotesize%
    $^1$Leiden University\\
    $^2$University of Amsterdam\\
    $^3$Utrecht University\\
    $^*$\href{mailto:e.bokanyi@uva.nl}{e.bokanyi@uva.nl}\\[2ex]
    \textit{version as of \today}
}}

\begin{document}

\maketitle

\onehalfspacing

\begin{abstract}

\noindent The dominance of online social media data as a source of population-scale social network studies has recently been challenged by networks constructed from government-curated register data. In this paper, we investigate how the two compare, focusing on aggregations of the Dutch online social network (OSN) Hyves and a register-based social network (RSN) of the Netherlands.
First and foremost, we find that the connectivity of the two population-scale networks is strikingly similar, especially between closeby municipalities, with more long-distance ties captured by the OSN.
This result holds when correcting for population density and geographical distance, notwithstanding that these two patterns appear to be the main drivers of connectivity.
Second, we show that the community structure of neither network follows strict administrative geographical delineations (e.g., provinces). Instead, communities appear to either center around large metropolitan areas or, outside of the country's most urbanized area, are comprised of large blocks of interdependent municipalities.
Interestingly, beyond population and distance-related patterns, communities also highlight the persistence of deeply rooted historical and sociocultural communities based on religion.
The results of this study suggest that both online social networks and register-based social networks are valuable resources for insights into the social network structure of an entire population.

\end{abstract}

\setlength{\parindent}{0em}
\setlength{\parskip}{0.8em}

\section{Introduction}

Until recently, population-scale social network analysis was typically done using data from online social networks (OSN) or mobile phone communication records. These digital data sources offer researchers unprecedented access to large amounts of information on human interactions and behavior \parencite{kumar2006structure, mislove2007measurement,blondel2015survey,eagle2009inferring}. Several works have attempted to understand the structural properties of global OSNs such as Facebook or Twitter \parencite{myers2014information, ugander_anatomy_2011}, as well as those of more localized ones, e.g., Hyves in the Netherlands \parencite{corten_composition_2012}, or iWiW in Hungary \parencite{lengyel2015geographies}. 
To understand how these social networks model the complex intricacies of the underlying societies, topological properties have been linked to various socio-economic outcomes. For example, economic connectedness of geographic areas is associated with upward social mobility \parencite{chetty_social_2022}, the abundance and diversity of connections is linked to economic prosperity \parencite{eagle2010network}, and inequality is reflected in more fragmented or closed network structures \parencite{toth_inequality_2021, kovacs_income-related_2023}. 

Recently, administrative government-curated records have become a novel population-scale  resource for register-based social networks (RSN). The use of administrative records is not entirely new; for example, employment has been widely used to analyze labor market outcomes \parencite{lyttelton_organizationally_2022,lyttelton_dual_2023,toth2022technology}. Uniquely, Statistics Netherlands recently combined multiple registers of people's connections of family, school, work, household, and next-door neighbors into a unique population-scale social network with multiple edge types \parencite{van_der_laan_person_2022_correct}. Such a register-based social network (RSN) models, typically for a well-delineated population, the social opportunity structure of people, and how this, for example, varies by age and different socio-economic variables such as income or education \parencite{bokanyi_anatomy_2023}. 
The structure of these networks has been shown to offer new insights into persistent social issues such as segregation \parencite{kazmina2024socioeconomic} or the intergroup connectivity of migrants and natives \parencite{soler2024contacts}.

Neither data sources (OSN nor RSN) have originally been designed for research; as such, they both present different opportunities and challenges. OSNs offer large-scale, automated data collection, often but certainly not always combined with self-reported data on people's demographic characteristics. However, the sample of both the nodes and the edges might not be representative. For one, this is because it is often unclear what exact social connections the edges represent. It can be difficult to differentiate bots from human agents, to find multiple profiles belonging to the same person, or to identify inactive or spurious connections \parencite{lazer_meaningful_2021,corten_composition_2012}. Because RSNs aggregate data from government-curated registers \parencite{van_der_laan_producing_2017_correct}, they offer legally defined high-quality data on nodes and edges. On the other hand, these edges only describe a so-called social opportunity structure \parencite{bokanyi_anatomy_2023, kazmina2024socioeconomic}, and we have no information on whether people actively use the connections.

So far, research cross-matching and consistently comparing OSN and RSN data is nonexistent. The ties in RSN data represent legally defined relationships - people are connected through \emph{formal ties} of kinship, work and school affiliations, and their registered address. Social tie formation for \emph{informal connections} in OSNs is usually explained based on concepts such as homophily \parencite{mcpherson2001birds} of, e.g., demographics or beliefs, triadic closure \parencite{asikainen2020cumulative}, or geographic distance between people \parencite{lambiotte2008geographical,liben-nowell2005geographic}. Each of these is expected to influence the probability that a tie exists between two people. \cite{vaneijk2010unequal} shows that a large share of informal ties come from current or former formal ties of people such as work, school, or family connections. The comparison of the two types of networks (OSN and RSN) in this paper thus attempts to advance the reconciliation of these two seemingly different social tie definitions.

In this work, we want to understand what influence the choice of population-scale data source has on network analysis research results, particularly on connectivity and community structure. To ensure the privacy of individuals~\parencite{de_jong_effect_2024}, we aggregate our datasets at the municipality level, and then ask a number of important questions related to how the data sources compare. Is the number of connections between municipalities similar between the two networks? What are the types of connections (e.g. family, work, or school) from RSNs that are best represented in OSNs? How do population size and geographical distance, i.e., factors known to affect social network connections \parencite{liben-nowell2005geographic, lambiotte2008geographical, krings2009urban, xu2022distance}, impact each of the two networks? Are meso-scale network structures, such as communities~\parencite{fortunato2010community, expert_uncovering_2011, fortunato2016community}, persistent across different data sources? And are these communities indicative of predefined administrative delineations, or do they reveal other patterns of group connectivity in the population-scale social network?

In this paper, we address the above questions using the unique combination of the Dutch online social network (OSN) Hyves of 6.2M nodes and 320M edges, and a register-based multilayer social network (RSN) of the entire population of the Netherlands of 16.6M nodes and 570M edges. 
We find that the number of connections between municipalities is strikingly similar in the two networks. 
Each type of connection, modeled as layers of the RSN (family, work, and school), uniquely contributes to this similarity. 
Comparing connectivity after removing the effects of population density and distance dependence reveals that while the local network structure is comparable, the OSN captures a larger number of distant connections. The community structure of both networks does not follow strict administrative geographical delineations but instead reveals deeply rooted sociocultural effects in both networks. 
In general, the findings presented in this paper show that both online social networks and register networks are useful for modeling the social network structure of a population.
This finding is important because across different countries, more and more population-scale register-based social networks are expected to be available for research in the future \parencite{magnani2022generation,savcisens2024using}.

\section{Results}

In this section, we first give a brief overview of the population-scale OSN and RSN datasets used. Then we provide two sets of empirical results.
The first pertains \emph{connectivity}, and compares the number of edges between municipalities in both networks, investigating how the different types of edges, that is, layers in the RSN, compare to the OSN. 
The second set of experiments dives into the \emph{community structure}, focusing on a comparison with administrative borders with the aim of understanding sociocultural aspects of the connectivity patterns. 
In both sets of experiments, we consider normalization of the connection strength by two common aspects known to influence connectivity: population size and geographical distance.

\subsection{Online and register-based population-scale social network datasets}

\begin{figure}[h!]
    \centering
    \includegraphics[width=1.2\textwidth,center]{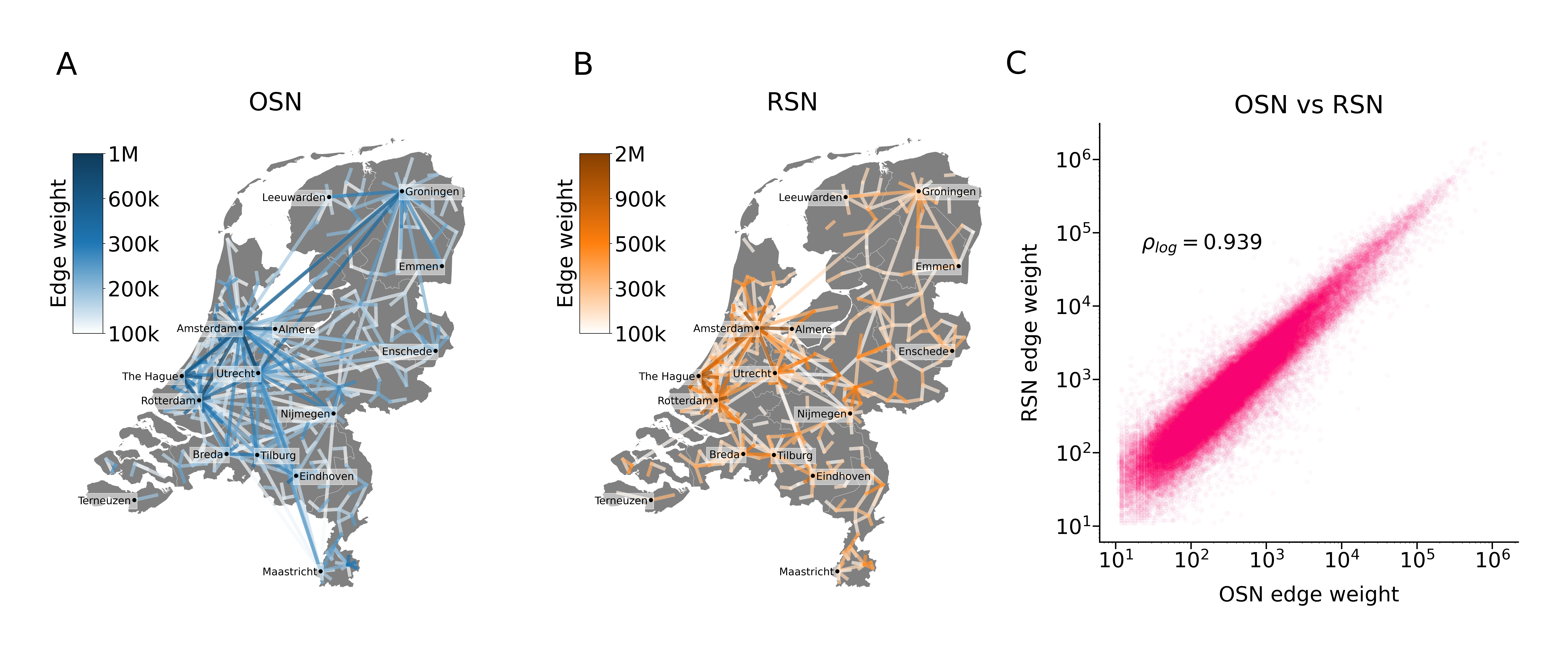}
    \caption{Spatial layout of the top 500 edges between municipalities in \textbf{(A)} the OSN  and \textbf{(B)} RSN. \textbf{(C)} Scatterplot of OSN and RSN edge weights between municipality pairs.}
    \label{fig:eszter-fig-1}
\end{figure}

We use a unique combination of two population-scale networks, each from 2010: the formerly highly popular Dutch online social network Hyves~\parencite{corten_composition_2012}, and a register-based social network constructed by Statistics Netherlands~\parencite{van_der_laan_person_2022_correct}.The Hyves dataset is an anonymized version of the service provider used in the study of \cite{corten_composition_2012}. The register-based social network was accessed and analyzed using the Remote Access environment of Statistics Netherlands. Hyves includes 6.2M users and more than 320M edges, and was the most popular online social networking site before the advent of Facebook.
The register-based social network (RSN) contains all 16.6M registered residents of The Netherlands as of 1 January 2010. The roughly 570M edges between the nodes include formal ties of current family, school, and work relationships sourced from administrative databases. These different types of edges constitute the layers of the RSN. The works of \cite{corten_composition_2012} and \cite{bokanyi_anatomy_2023} contain more details on the topological properties of both networks.

Direct person-level matching of the two networks is legally and technically impossible due to privacy and record linkage limitations. Thus, we aggregated the two networks into the 431 municipalities of the Netherlands in the year 2010. Weighted edges between the municipality pairs denote the number of ties between people in the given network. Figure~\ref{fig:eszter-fig-1}A and \ref{fig:eszter-fig-1}B depict the 500 edges with the largest weight in both networks. 
At the endpoints of these strong edges, there are major cities with large populations, but some of them are in rural areas. 
For more details on the two networks and the construction of the municipality-level aggregation, see the \nameref{methods-and-data} section. 

\subsection{Connectivity in the OSN and the RSN}

In the following set of analyses, we seek to understand the structure of the OSN and the RSN in terms of connectivity.
We do so for the weighted networks themselves, as well as for versions of the network normalized by population density and geographical distance (see \nameref{methods-and-data} for details). 
For each of the three variants of the network, we then look at the similarity of different individual layers of the RSN and the OSN. 

First, we compare the edge weight of the municipality pairs in the OSN and in the RSN in Figure~\ref{fig:eszter-fig-1}C. The Pearson correlation of the logarithm of the weights is 0.939. This suggests a high micro-level similarity between the two networks. However, this similarity could be driven by factors that are known to influence the number of connections, such as population size and geographical distance between municipalities.
In the following, we break down how similar individual RSN layers (e.g., family, school, or work) are to the OSN, how normalizing by population and distance influences the similarity.

\begin{figure}[h!]
    \centering
    \includegraphics[width=\textwidth]{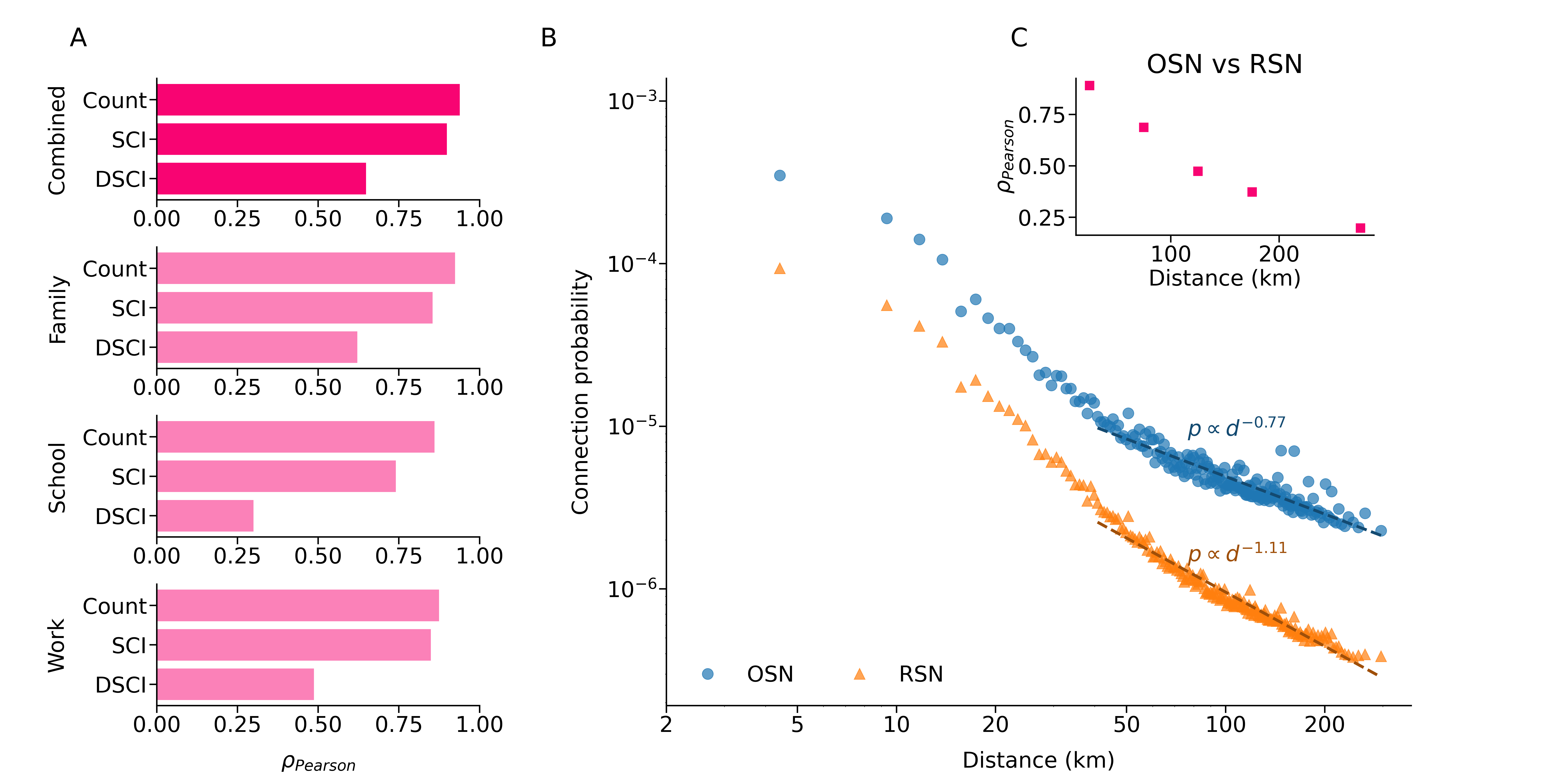}
    \caption{    
    \textbf{(A)} Pearson correlation similarity of OSN and RSN edge weights using all RSN layers (Combined), and Family, Work, and School, each for different edge weighting strategies (Count, SCI an DSCI).
    \textbf{(B)} Distance dependence of connection probabilities in the OSN and RSN \textbf{(C)} Correlation of DSCI edge weights for different distance ranges.}
    \label{fig:eszter-fig-2}
\end{figure}

Figure~~\ref{fig:eszter-fig-2}A shows the similarity between the OSN and the RSN for all possible connection types of the RSN (Combined), and the family, school, and work edge types. The correlations are calculated for the three different edge weighting strategies: the Count capturing the number of connections between two municipalities, the population-normalized SCI, and the distance-normalized DSCI. The first correlation of the top row shows the Pearson correlation of 0.939 from Figure~\ref{fig:eszter-fig-1}C. 
If we restrict the RSN to a single layer type, we get lower correlation values: 0.924 for the family, 0.860 for the school, and 0.874 for the work layer. Thus, combining all available RSN edge types (layers) gives the highest similarity to the OSN.

Population size is an important driving factor for connectivity, as larger municipalities naturally have more connections. The Social Connectedness Index (SCI) metric of \cite{bailey_social_2018} aims to correct for this dependence by normalizing the edge weights by the population size of the edge endpoints. Focusing on the results for SCI in Figure~\ref{fig:eszter-fig-2}A, we observe slightly lower but still high similarities compared to the plain edge weights: 0.899 for the combined layers, 0.854 for the family, 0.741 for the school, and 0.849 for the work layers. This suggests that the edge weights of the two networks are similar beyond population distribution patterns. Again, combining layers gives the highest similarity.

These correlations might still be partially driven by distance dependence, namely that closeby places are more likely to be connected. This relationship is often formalized as a gravity law \parencite{lambiotte2008geographical}, in which the connection probability between areas has a power-law dependence on the Euclidean distance with a negative exponent. Looking at Figure~\ref{fig:eszter-fig-2}B, we can observe this power law in the OSN and in the RSN with tail exponents -0.77 and -1.11, respectively. The OSN has a larger (negative) power-law exponent which indicates more large-distance connections that are not as sensitive to distance as in the RSN. It should be noted that the overall higher probability of the OSN edges at all distances can be attributed to normalizing the probabilities by user counts instead of population size.

To account for the distance dependence of the connection probability, we propose DSCI: a distance-aware SCI metric that beyond population, is also normalized by the expected SCI for a certain distance. Figure~\ref{fig:eszter-fig-2}A shows that DSCI correlations between the OSN and RSN drop significantly compared to the plain SCI measure: the similarity of the OSN and the combined layers of the RSN is 0.649, the family layer is 0.621, the school layer is 0.300 and the work layer is 0.487. This reflects the influence of factors in social tie formation that go beyond population size or spatial distance.

Because the RSN has more large-distance connections, we investigate how the correlation changes if we restrict the calculations to edges of different distances in Figure~\ref{fig:eszter-fig-2}C. We find that indeed the correlation between DSCI edge weights decreases as the distance increases.

\subsection{Community structure}

In this section, we explore the community structure of the OSN and the RSN. We particularly investigate how closely these align between the two networks and with existing administrative boundaries, and what other connectivity patterns can be revealed by the obtained community structure.

\begin{figure}[p]
  \centering
  \includegraphics[width=\textwidth]{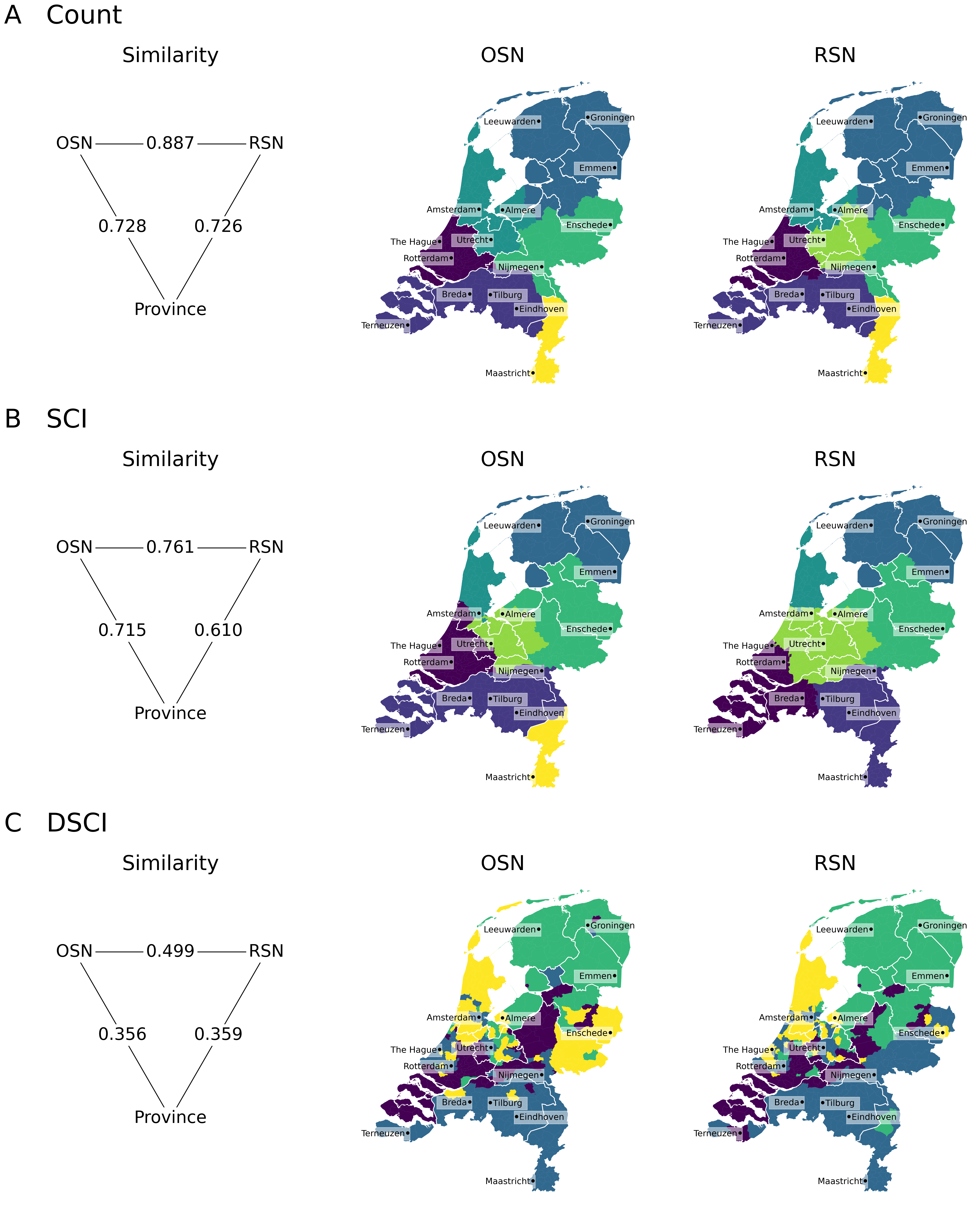}
  \caption{Community structure similarity of the online social network (OSN), the register-based social network (RSN), and administrative borders (Provinces), measured in terms of Adjusted Mutual Information (AMI) for \textbf{(A)} the plain weighted network, \textbf{(B)} weights normalized by population density (SCI) and  \textbf{(C)} population density and distance (DSCI).}
  \label{fig:eszter-fig-3}
\end{figure}

We use the well-known Louvain modularity optimization algorithm~\parencite{blondel_fast_2008} combined with consensus clustering~\parencite{kwak_consistent_2011,mandaglio_consensus_2018} to minimize the effects of randomness inherently present in such algorithms. For further explanation, see the \nameref{methods-and-data} section. The pairwise similarities of the communities found are calculated using the adjusted mutual information (AMI) metric. An AMI value of 1 means identical partitions and 0 means that the partitions are only as similar as expected due to random chance.

In the triangles of Figure~\ref{fig:eszter-fig-3}A-C, we can see the AMI similarity scores comparing partitionings of the OSN and the RSN. Figure~\ref{fig:eszter-fig-3}A is based on the raw edge weights, Figure~\ref{fig:eszter-fig-3}B uses the SCI, Figure~\ref{fig:eszter-fig-3}C the DSCI metric. The similarity between the communities of the two networks is high for the counts and the SCI metric: 0.887 and 0.761, respectively. The country maps of Figures \ref{fig:eszter-fig-3}A and \ref{fig:eszter-fig-3}B displaying the community detection results show spatially contingent community structures. This highlights a localized geographical preference in group formation.

If we normalize edge weights for distance as well, i.e., look at DSCI in Figure~\ref{fig:eszter-fig-3}C, we get a lower similarity, 0.449, between the two networks. The visual representation of the results on the map suggests that the communities are no longer spatially contiguous. Despite the slightly lower similarity values, we find one specific commmunity in both networks that spans from the southwest to the northeast. This roughly follows the area of the so-called Bible Belt of the Netherlands, a set of regions that form a distinct sociocultural unit with high shares of religious adherence and conservative voters \parencite{exalto2019strong, rellstab2023gender}.

The bottom two sides of the triangles in the leftmost column of Figure~\ref{fig:eszter-fig-3}A-C indicate similarity of our resulting partitions to pre-established administrative boundaries, in our case, the subdivision of the Netherlands into 12 provinces. We highlight the borders of these provinces in Figure~\ref{fig:eszter-fig-3}A-C.
The first two edge weighting strategies (Count and SCI) provide relatively similar tendencies to province borders with AMI scores for both networks between 0.728, 0.726, 0.715, and 0.610. We can see that some clusters largely follow province borders, with only small deviations, whereas other clusters span multiple provinces. However, using DSCI, we can observe a notable difference to the province borders. This further supports the assumption that DSCI weighting can capture important socio-economic similarities other than the population and proximity bias.

\section{Discussion}

It is well-known in the social network analysis literature that each source of social network data comes with its particular biases, as well as data completeness and data quality issues~\parencite{lazer_meaningful_2021}. We have analyzed the similarities between two Dutch population-scale networks of different sources: Hyves, an online social network (OSN), and a register-based social network of Statistics Netherlands (RSN), each aggregated to the municipality level. 

On the level of municipality pairs, we found that the two networks are very similar in terms of connectivity. Although there is a relatively high correspondence in the edge weights even when only using a single layer from the RSN as family, work, or school; the RSN is most similar to the OSN when including all available edge types, thus when combining multiple contexts of life in the construction of the register-based network. This highlights that behind the 'friend' edges of online social networks, there may be multiple different mechanisms at play when users establish connections. It also suggests that the more relationships a register-based network can incorporate, the better it reflects the social opportunity structures of people, and the better we can generalize results obtained from the RSN to other datasets. This result is also in line with the literature that suggests that a large share of informal relationships are based on various forms of current or former family, school, and work connections of people \parencite{vaneijk2010unequal}. Thus, even though a register-based network does not capture informal connections by definition, a superposition of various formal connections performs well when modeling an aggregated social network of a whole country.

The similarity in edge weights can be driven by the fact that the same main factors of edge formation drive the connectivity in both networks. Therefore, we performed two normalization approaches: we calculated the SCI, a population-normalized version of the edge weights, and the DSCI, the deviation from the expected SCI at a given geographical distance. We find that the Pearson correlation of the edge weights decreases if we apply the normalization, but it remains at a relatively high value of 0.6 even for DSCI. Thus, connectivity reflects patterns of tie formation even after accounting for the population density and distance effects. The remaining similarity most likely captures socio-cultural homophily and geographical or economic constraints, footprints which are in the micro-level structure of both types of networks. The fact that the correlation of DSCI decreases with increasing distance highlights that local structures are strikingly similar in OSNs and RSNs. However, long-distance connections are both more numerous and more random in online social networks. It is important to consider these findings when interpreting the results from spatial networks on a large geographical scale.

Reported findings on the community structure suggest that the structure of the two networks is also similar at the meso-scale.
We find the least correspondence between the OSN and RSN communities when using normalization for both distance and population. However, in this latter case, a remarkable community which is not geographically contiguous, appears in both the OSN and RSN. It includes most of the so-called Dutch Bible Belt, which is a set of regions that form a distinct sociocultural unit with high shares of religious adherence and conservative voters \parencite{exalto2019strong, rellstab2023gender}. Thus, distance- and population-aware communities uncover socioculturally similar regions in both networks in a similar way to \cite{expert_uncovering_2011}. If we compare communities with the administrative boundaries of Dutch provinces, we find that distance-unaware communities somewhat correspond to administrative boundaries. However, there are notable differences, such as the northern part of the province Flevoland attaching to Overijssel. This might reflect that economic and infrastructural constraints matter more than administrative division in this case, since the Northern parts of Flevoland are infrastructurally well-connected to the neighboring province. When using DSCI, there is very little agreement between network communities with province boundaries. Hence, socioeconomic policy making on certain topics such as labor markets, infrastructure investments, or formal care systems might be better based on community clusters rather than provincial boundaries. 
 
The differences in the micro- and meso-scale structures of the OSN and RSN highlighted throughout this paper might originate from the different underlying link generation strategies in the two networks. In the RSN, family links represent persistent ties, but work and school relationships only reflect the situation of the current year. OSNs better capture the fact that some relationships are always retained from former schools or workplaces, even if sometimes in a different context such as a close friendship. Aggregating RSN links over time and comparing them with the OSN structure could provide further insight into this matter. On the other hand, OSN links might reflect connections of very different strengths, ranging from close family to distant past aquaintances. This can partly explain the differences in the structure of large-distance connections between the RSN and the OSN. RSNs miss out on important informal relationships such as church or leisure groups. It is important to note that the person-level degree distributions of the two networks are different. In the RSN, most people have a typical number of connections \parencite{bokanyi_anatomy_2023}. 
In Hyves, there are many low-degree nodes and fewer high-degree nodes \parencite{corten_composition_2012}. 
Interestingly, we observe the structural similarity of the two municipality-level networks despite these differences.

\section{Conclusion}

In this work, we provided an in-depth comparison of the population-scale network structure of an online social network and a register-based social network in the Netherlands. 
We observed similarities between the micro and meso-level structure of the two networks despite the OSN containing self-reported friendship ties, and the RSN being based on legal definitions of kinship and formal affiliations such as work and school. We showed
that the two networks are strikingly similar when comparing their connectivity;
that combining all available RSN layers (family, school, work) results in the highest similarity;
and that similarity remains relatively high even after accounting for population and spatial distance patterns, especially for local edges.

By analyzing communities of the two networks with different edge weighting strategies, we showed
that the networks have similar community structures using all three edge weighting strategies;
and that detected communities do not closely follow pre-established administrative borders, especially when accounting for population and distance patterns. However, the latter method uncovers a socioculturally tightly knit community that corresponds to the Dutch Bible Belt.

In summary, we expect researchers to draw similar conclusions based on register-based social networks and online social networks, especially for short-distance connections. Both data sources are useful for modeling the social network structure of a whole population, and the more edge types a register-based network contains, the better the comparability.

\section{Data and Methods} \label{methods-and-data}

In this section, we first present Hyves and the register-based social network of Statistics Netherlands, the OSN and RSN datasets, and their aggregation into Dutch municipalities, followed by a detailed description of the RSN layers. Then, we introduce our notation and describe the methods for normalizing edge weights. Lastly, we outline the process of identifying communities.

\subsection{Social network datasets}

\textbf{OSN}. The Hyves online social network was an online social media platform in the Netherlands~\parencite{corten_composition_2012} before the advent of Facebook. The dataset represents the late 2009 state of the network which during its peak period contained 10M people, covering up to 60\% of the population of the Netherlands. The network represents supposed friendship connections between its registered users. There are 6.2M users with a self-reported place of residence at a municipality-level resolution, with 320M edges between them. We excluded users flagged as celebrities from our analysis. 

The self-reported municipality names in the data were provided by users and therefore were prone to different errors. \cite{norbutas2018network} cleaned and aggregated place names at the municipality level even if users gave different administrative units, such as neighborhoods, as their place of residence. The municipalities were matched to the official list of Statistics Netherlands as of 2009. We use this cleaned and matched municipality dataset, and refer the reader to \cite{norbutas2018network} for more details on data processing.

\textbf{RSN}. The register-based social network (RSN) is compiled from official records of and by Statistics Netherlands (CBS)~\parencite{van_der_laan_producing_2017_correct,van_der_laan_person_2022_correct}. In this network, the nodes are all 16.6M residents of the Netherlands in 2010. The almost 800M edges are organized in several layers representing various contexts of life comprising current family, school, work, neighbor and household relations. Each person's place of residence (municipality) is known. Only family, school and work connections are meaningful when considering inter-municipality connections, thus only these 570M edges are retained when aggregating connections at the municipality level.

\textit{Family} Family connections are derived from official parent-child and partner relations. The partner relations are derived from marriage registers, tax declarations, and household registers. From the above two source datasets, other family ties such as grandparents, grandchildren, siblings, aunts, uncles, cousins, nieces, and nephews are inferred. Step and in-law relationships are also included.

\textit{School} School connections are aggregated from various official educational agencies containing five levels of education: elementary, secondary, secondary special, vocational, and higher. People have a school tie if they go to the same school, year, location, and type of education. University and other higher education students are further distinguished by study programmes.

\textit{Work} Work connections contain links between people working for the same employer of their major source of income. If a company has less than 100 employees, all of them are connected to each other. Otherwise, a person is connected only to the 100 co-workers closest to their residence.

We introduce the intuition and notion of the aggregated network of municipalities created from the person network. As direct person matching between the two networks is infeasible, we aggregate the networks such that nodes are the 431 municipalities of the Netherlands in 2010, and weighted edges between municipality pairs count the number of ties between people in the municipalities that the link connects. There were a few municipality merges in The Netherlands from 2009 to 2010 that we applied to the network dataset as well. We obtain this aggregated network for the OSN and every relevant layer in the RSN (family, school, and work), as well as for a combination of these layers. In the combined layers, an edge exists between two people when at least an edge exists in any of the three layers. If multiple edges run between people in the base layers, we count it as a single edge in the aggregated layer.

\subsection{Preliminaries}

We introduce the notion of the aggregated multilayer graph. The notation is based on \parencite{bokanyi_anatomy_2023}. 
We represent a person-level network as $G_p=(V_p, E_p, L)$, where $V_p$ is the set of nodes respresenting people. In our case, the residents of the Netherlands in 2010 consisted of $|V_p|=n_p=16.6$M people. The set of undirected edges running between these nodes can be described as 

\begin{equation}
    E_p \subseteq \{ (\{u,v\},\ell) \, : \, u,v\in V_p, \, u \neq v, \, \ell \in L \},
\end{equation}
such that $L$ is the set of possible layers. 
In this setting, we can represent the network $G_p$ using personal level binary adjacency tensor $(A_p)_{u,v,\ell}$. An entry $a_{u,v,\ell}$ of this matrix is 1 if and only if an edge runs between persons $u,v \in V_p$ in layer $l \in L$, and 0 otherwise. 

We define $G=(V, E, L)$ as a multilayer graph. In this case, $V$ is the set of municipalities, $n=|V|=$ 431 is the number of nodes. The set of undirected edges is
\begin{equation}
    E \subseteq \{ (\{u,v\},\ell, w) \, : \, u,v\in V, \, u \neq v, \, w \in \mathbb R , \, \ell \in L \},
\end{equation}
such that $L$ is the set of the possible layers and $w$ is the strength of the connection between the two municipalities, for which we propose three different weighting schemes in the \nameref{weightingschemes} section below. We can represent the edges $E$ using an adjacency matrix $A_{u,v,\ell}$ that counts half-edges that run between $u,v \in V$ in the layer $\ell\in L$.

We can relate the two representations as follows. We can represent the place of residence using a binary affiliation matrix $B$ of shape $n_p \times n$. An entry in this matrix is 1 if $u\in V_p$ is affiliated to $v \in V$. With the help of this representation, we can calculate 
\begin{equation}
    A_{u,v,\ell} = B^T (A_p)_{u,v,\ell} B.
\end{equation}
Here, $(.)^T$ represents matrix transposition.

Person-level edges can also originate and end in the same municipality. This is represented by weighted self-edges in the aggregated graph. The above equation would count person-level edges twice within the same municipality. However, we dropped all of the self-edges when running our experiments.

\subsection{Weighting schemes}
\label{weightingschemes} 

We compare the strengths of the connections between municipality pairs in the RSN and the OSN by comparing edge weights corresponding to the same municipality pair. We incorporate population and distance into the weighting scheme as follows.

\textbf{Population corrected weighting.} We use a metric that not only counts connections between areas but also takes into account that larger population areas typically have more connections inspired by~\parencite{bailey_social_2018}. Within a layer, the metric between $i,j \in V$ is formulated as
\begin{equation} \label{eq:sci}
    SCI_{ij} = \frac{\text{Connections}_{ij}}{\text{Possible connections}_{ij}}.
\end{equation}
If $i \neq j$, then the number of possible connections is Population$_i \times $ Population$_j$. Otherwise, it is equal to Population$_i \times ($Population$_j -1)$. In the case of the RSN, the population is the number of inhabitants. In the case of the OSN, population is the number of users that have self-reported the municipality as location in their profiles.

\textbf{Population and distance corrected weighting}. It is widely known that distance is an important factor when forming connections \parencite{lambiotte2008geographical}. A power-law distribution often models this dependency which is often called the gravity law. In our context, we use a model-free metric inspired by~\parencite{expert_uncovering_2011}. To measure the distances of municipalities, we calculate the distances of the centroids of the municipalities using the Euclidean distance metric.

The proposed distance-aware social connectivity index (DSCI) is given by
\begin{equation}
    DSCI_{ij,D} = \frac{SCI_{ij}}{\mathbb E \left[ SCI | D \right] },
\end{equation}
where $D$ denotes a certain spatial distance and $\mathbb E[ \cdot ]$ denotes the expected value of a variable. We approximate this value by creating 200 bins that contain an equal number (464) of municipality pairs between 0 and 360 km.

\subsection{Community detection}

Community detection \parencite{newman_modularity_2006,girvan_community_2002} is a way to identify groups of nodes in a network that form tightly knit subunits that are more loosely connected with other subunits. We use this to investigate the meso-level structure in our networks.  We perform community detection based on the Louvain method, which accounts for edge weights. This allows us to investigate the three community structures resulting from our three edge weighting strategies: the number of connections between municipalities, the SCI weights, and the DSCI weights. We set the resolution parameter to 1. We use the \t{Python} package \t{networkx} \parencite{hagberg_exploring_2008}.

It is well-known that community detection algorithms involve a degree of randomization. This can be accounted for by using consensus clustering \parencite{lancichinetti2012consensus}. 
In our experiments, we use 1) 1000 iterations of the Louvain algorithm to 2) create a new network based on the number of times the nodes belonged to the same community. Then, we go to step 1) and repeat until convergence. In our experiments, it took 3 iterations until all node pairs distinctively belonged to the same community. The results can also be regarded as a partitioning of the node set $V$ into $R$ non-empty partitions $U = \{U_1, U_2, \dots, U_R\}$, where $U_i\cap U_j = \{\}$ for any $i\neq j,\ i,j\in\{1,2,\dots,R\}$, so the partitions are pairwise disjoint, and $\cup_{i=1}^R U_i = V$.

We use the Adjusted Mutual Information (AMI) metric of \cite{vinh2009information} to compare the partitions we get from the concensus clustering on the different networks and edge weighting strategies, and also to compare the network partitions to the province borders of the Netherlands which is in essence an administrative partitioning. If we have two different partitionings, $U=\{U_1, U_2, \dots, U_R\}$ of $R$ partitions, and  $T = \{T_1, T_2, \dots, T_C\}$ of $C$ partitions
, then the Adjusted Mutual Information is:
\begin{equation}
    AMI(U,T) = \frac{
            MI(U,T) - \mathbb{E}\left\{MI(U,T)\right\}
        }{
            \mathrm{avg}\,\left\{H(U),H(T)\right\} - 
            \mathbb{E}\left\{MI(U,T)\right\}
        }.
\end{equation}

In the above equation, $MI$ stands for Mutual Information, which if calculated as 
\begin{equation}
    MI(U,T) = \sum_{i=1}^R\sum_{j=1}^C P_{UT}(i,j)\log\frac{P_{UT}(i,j)}{P_U(i)P_T(j)},
\end{equation}

where $P_{UT}(i,j) = \frac{|U_i\cap T_j|}{|V|}$ is the probability that a node belongs to partition $i$ in $U$, and partition $j$ in $T$, and $P_U(i) = \frac{|U_i|}{|V|}$ is the probability that a node belongs to partition $i$ in $U$, and $P_T(j) = \frac{|T_j|}{|V|}$ is the probability that a node belongs to partition $j$ in $T$. $\mathbb{E}$ denotes the expected value of the Mutual Information, for details on its calculations, we refer the reader to \cite{vinh2009information}. The expected MI terms normalize this score to reflect that two random partitionings can also have similarity by chance.

$H$ stands for the entropy associated with a partitioning $U$:
\begin{equation}
    H(U) = \sum_i^R -P_U(i)\log P_U(i).
\end{equation}

We use the implementation of the \texttt{scikit-learn} \parencite{pedregosa2011scikitlearn} Python package (see \texttt{sklearn.metrics.adjusted\_mutual\_info\_score}) for the calculations.

\subsection*{Acknowledgements} 

We would like thank the POPNET team~(\url{www.popnet.io}) for helpful suggestions and discussions. The POPNET project has been funded by Platform Digitale Infrastructuur Social Sciences and Humanities~(\url{www.pdi-ssh.nl}). 

\subsection*{Data availability}

Results are based on calculations by Márton Menyhért (Leiden University) and Eszter Bokányi (University of Amsterdam) as part of the POPNET project~(\url{www.popnet.io}) using non-public microdata from Statistics Netherlands. Under certain conditions, these microdata are accessible for statistical and scientific research. For further information , contact the corresponding author and: \href{mailto:microdata@cbs.nl}{microdata@cbs.nl}.

\printbibliography

\end{document}